\documentclass[conference]{IEEEtran}
\usepackage{url,cite,amssymb,amsmath,nccmath,subfigure,multirow}
\usepackage[pdftex]{graphicx}
\usepackage{flushend}
\usepackage{color}
\begin{document}

\title{Software-Defined Cloud Computing: Architectural Elements and Open Challenges}

\author{\IEEEauthorblockN{Rajkumar Buyya\IEEEauthorrefmark{1},
Rodrigo N. Calheiros\IEEEauthorrefmark{1},
Jungmin Son\IEEEauthorrefmark{1},
Amir Vahid Dastjerdi\IEEEauthorrefmark{1}, and
Young Yoon\IEEEauthorrefmark{2}\\[5pt]}
\IEEEauthorblockA{\IEEEauthorrefmark{1}\textbf{Clou}d Computing and \textbf{D}istributed \textbf{S}ystems (CLOUDS) Laboratory\\
Department of Computing and Information Systems\\
The University of Melbourne, Australia\\
Email: \{rnc@, jungmins@student, rbuyya@\}unimelb.edu.au\\[6pt]}
\IEEEauthorblockA{\IEEEauthorrefmark{2}Samsung Electronics, Korea\\
Email: young77.yoon@samsung.com}}

\maketitle

\begin{abstract}
The variety of existing cloud services creates a challenge for service providers to enforce reasonable Software Level Agreements (SLA) stating the Quality of Service (QoS) and penalties in case QoS is not achieved. To avoid such penalties at the same time that the infrastructure operates with minimum energy and resource wastage, constant monitoring and adaptation of the infrastructure is needed. We refer to Software-Defined Cloud Computing, or simply Software-Defined Clouds (SDC), as an approach for automating the process of optimal cloud configuration by extending virtualization concept to all resources in a data center. An SDC enables easy reconfiguration and adaptation of physical resources in a cloud infrastructure, to better accommodate the demand on QoS through a software that can describe and manage various aspects comprising the cloud environment. In this paper, we present an architecture for SDCs on data centers with emphasis on mobile cloud applications. We present an evaluation, showcasing the potential of SDC in two use cases---QoS-aware bandwidth allocation and bandwidth-aware, energy-efficient VM placement---and discuss the research challenges and opportunities in this emerging area.
\end{abstract}

\section{Introduction}

Cloud computing \cite{buyya:09} evolved as a successful utility computing paradigm for Information and Communication Technology (ICT) resources delivery as a service over the Internet. The adoption of cloud computing spans across industry, governments, and academia alike.

With the growing adoption of cloud, the number of cloud providers and services also has increased. Hundreds of providers are offering services in either of the three service models---Software as a Service (SaaS), Platform as a Service (PaaS), and Infrastructure as a Service (IaaS). Furthermore, there is a wide array of products offered by each provider for each service model, and each product can be configured with many different parameters.

This variety of services creates a challenge for service providers to enforce reasonable Service Level Agreements (SLAs) stating the Quality of Service (QoS) and penalties in case it is not achieved. Since SLA establishment is legally required for compliance and has potential impact on revenue, meeting these SLAs is a primary concern of service providers.

One elementary approach to guarantee SLA compliance is over-provisioning of resources for a given service: by committing more resources than the service actually requires, a provider can meet acceptable level of QoS even in the presence of failures of one or more resources. Nevertheless this approach is not economically viable, i.e., over-provisioning increases the cost of running the service.


An alternative approach involves constant monitoring and adaptation of the infrastructure, in such a way that the minimum amount of resources required to meet the SLA (promised to a particular group of users) is committed to the service, and constant monitoring guarantees that whenever resources are under-performing, an appropriate corrective action is taken. Corrective actions include network reconfiguration (so delays can be reduced or bandwidth between resources can be increased), virtual machines (VMs) migration or reconfiguration, and storage commitment. Manual solutions cannot enable such actions to be taken in able time to mitigate the effect of under-performing resources in services violating their SLAs, because they require intervention of operators that need to understand the cause of under-performance and then react to it. Automatic approaches, on the other hand, can carry out the same activities more effectively, reducing the time to action.

An approach for creating effective cloud environments by extending virtualization notion to all resources (including compute, storage, and networks) of a data center is Software-Defined Cloud Computing, or simply Software-Defined Clouds (SDC). SDCs enable easy reconfiguration and adaptation of physical resources in a cloud infrastructure to better accommodate the demand on QoS. An SDC allows each aspect that comprises the cloud environment---from the networking between virtual machines to SLAs of hosted applications---to be managed via software. This reduces the complexity related to configuring and operating the infrastructure, which would ease the management of cloud infrastructure that tends to be very large scale, commonly composed of thousands of servers and network elements supporting tens of thousands of virtual machines, virtual networks, and applications. Furthermore, the awareness of the huge carbon and energy footprint of cloud  data centers has become a crucial cloud system design criteria. Thus, every aspect of SDCs needs to operate efficiently in terms of energy consumption.

In this paper, we propose an architecture for SDCs targeting Web, compute-intensive, data-intensive, and mobile cloud applications. The proposed architecture is built upon recent advances in many areas: server virtualization~\cite{barham:03}, software-defined networks (SDNs)~\cite{sdnreport}, software-defined middlebox networking~\cite{gember:12}, and network virtualization~\cite{koponen:14}. This paper describes how these disparate concepts can be combined to deliver SDCs, discusses challenges arising from the emerging concept of SDC, and presents some early empirical insights into the new concept.

The rest of this paper is organized as follows. Section~\ref{sec:background} discusses the concepts and technologies that are the foundation for SDCs. Section~\ref{sec:related} presents related works. Section~\ref{sec:scenario} introduces the application scenario we explore in the paper, as well as requirements for SDCs. Section~\ref{sec:architecture} introduces the proposed architecture and its components in detail. Section~\ref{sec:evaluation} discusses the evaluation of the proposed architecture. Section~\ref{sec:challenges} discusses open challenges and opportunities in the emerging field of SDCs. Finally, Section~\ref{sec:conclusion} concludes the paper.

\section{Background} \label{sec:background}

Software-Defined Cloud Computing is enabled by a number of concepts and technologies that we briefly introduce in this section.

\subsection{System Virtualization}

System virtualization~\cite{smith:05} enables multiple operating systems to run in isolation on a single host. The technology was first introduced in the 60's with the objective of increasing the utilization of mainframes~\cite{goldberg:74} and, after a hiatus during the emergence of the personal computing era, gained traction again in the 2000's when PCs  became powerful enough to accommodate multiple operating systems~\cite{devine:02,barham:03}.

Isolation among OSs is enabled by a special software called \emph{hypervisor}, which also coordinates the resource sharing among VMs. The hypervisor is installed on top of the bare hardware and it is loaded on server initialization. It then provides management capabilities that enable client operating systems (referred to as \emph{guests}) to be started, paused, or destroyed on the server, as well as to experience variation in the amount of resources assigned to each machine (e.g., RAM). 

Each guest runs one or more virtual NICs that are connected to a virtual switch~\cite{jain:13}. Different approaches can be taken to connect the virtual switch to the physical network, varying from connecting it to the physical NIC of the server to connecting it directly to a physical switch~\cite{jain:13}.

System virtualization is a building block for other technologies of relevance for this work, such as cloud computing and Software-Defined Middleboxes Networking, which are explored later in this section.

\subsection{Cloud Computing}

Cloud computing~\cite{mell:11} is a paradigm that enables the acquisition of computational resources on demand in a pay-per-use model. Consumers of cloud computing resources can provision their own resources when they are  needed, and can release them quickly once they are not required. Resources are kept in large pools, giving the consumers the notion of an infinite amount of available resources. Resource utilization is metered by the cloud service provider.

There are three service models typically associated with cloud computing. Software as a Service (SaaS), offers to consumers ready to use applications. Thus, consumers neither need to install the application on their own infrastructure nor need to acquire licenses for the software. The cost of licensing (if any) is included in the hourly (or monthly) cost of the software service. Platform as a Service (PaaS) is a model that offers platforms where applications can be easily deployed. Consumers of PaaS services do not need to handle any intricacy of the underlying platforms, and deployment of whole web application stacks is as simple as pushing code to the system hosted by the service provider. Finally, Infrastructure as a Service (IaaS) offers virtual machines and other low level features for consumers, which are also responsible for managing the operating system on the VM and the entire software stack installed on it.

In this paper, we focus on IaaS when we refer to cloud services, as it offers the biggest flexibility to consumers and also can be a building block for the other service models.

\subsection{Mobile Cloud Computing}

Mobile cloud computing deals with the problem of enhancing the capacity of mobile devices by offloading resource-intensive applications to components external to the device itself~\cite{fernando:13}. Such external devices can be cloud servers~\cite{abolfazli:14}, other mobile devices~\cite{fernando:11}, or \emph{cloudlets} (powerful servers that are spread around the mobile's coverage area to provide extra capacity or to offload workload to the cloud)~\cite{satyanarayanan:09}. The motivation for offloading CPU-intensive tasks is to enable mobile devices to perform activities that otherwise would not be possible because of the limitations in resources on the device (e.g., energy, CPU, network bandwidth)~\cite{fernando:13}.

In this paper, we are particularly interested in the approach of offloading of tasks to a cloud data center. This reason is because other mobile devices in the same network are subject to a similar restriction in resources than the first device, and thus outsourcing the load to another device does not solve the problem of limited capacity---it just transfers the problem to someone else's device. Besides that, mobile device owners might have concerns about privacy and security of the data on their devices, which may discourage them to share their resources with unknown third parties. Finally, we chose the cloud approach over the cloudlet approach because it enables the vast amount of resources from existing data centers to be promptly leveraged for supporting mobile applications and enabling rapid elasticity of applications using a huge pool of resources.

\subsection{Software-Defined Networks}

Software-Defined Network (SDN) is a technology that proposes the isolation between the control and forward planes from networks. This enables higher level software to configure the control plane of network packets on demand, enabling the infrastructure to quickly adapt to new application requirements~\cite{sdnreport}. This is achieved with a new control layer between applications and the infrastructure that receives instructions from applications and emits configuration actions to the equipment in the infrastructure layer.

Such a new layer performs operations that previously were ingrained in the routing hardware. Prior to the development of this new layer, configuration of routing hardware was labor-intensive and error-prone, as it would require manual access to each piece of hardware.

Network devices are still responsible for the forwarding plane, where data packets from application transit from source to destination hosts. This happens with consideration for routing and priorities configured by the control layer to meet the demand faced by the network.

\subsection{Software-Defined Middleboxes Networking}

Middleboxes comprise of networking elements that perform specialized tasks, including security, load balancing, network address translation, and performance improvement. Each of these specialized operations is implemented in a hardware device and then is inserted in the network. Some of these operations can be CPU-intensive and may process a vast number of packets arriving at the devices. A recent trend in this area is the virtualization of such middleboxes, either via outsourcing such services to the cloud~\cite{sherry:12} or by the application of techniques similar to those from SDNs~\cite{qazi:13,sherry:12}.

Regarding the latter approach, it requires the middlebox functionality to be implemented in a virtual machine and then having this VM deployed and executed in the network. Given the CPU-intensive nature of middleboxes, virtualization of such services consume CPU resources that are also used by the virtual machines hosting the cloud applications. Therefore, it adds complexity to the general management of the virtualized system. Besides this issue, software-defined middleboxes networking shares some benefits of SDNs, including centralized management of components.

\subsection{Network Virtualization}

Network virtualization  enables multiple virtual networks to be established over one or more physical networks~\cite{chowdhury:10}. It has been widely adopted for enabling features such as virtual local area networks (VLANs), virtual private networks (VPCs), and overlay networks~\cite{chowdhury:10}. Recent proposals focused on aspects of network virtualization that target multi-tenant data centers~\cite{koponen:14}, and thus they are relevant to cloud data centers.

The focus of such recent works is on ensuring that virtual machines on the same physical network (or even on the same host) are securely isolated at the network level from machines that share the same physical networking resources.

\section{Related Work} \label{sec:related}

Grandl et al.~\cite{grandl:13} proposed a system called Harmony that manages aspects of compute, storage, and networks for Software-Defined Clouds. Our architecture considers these element and also considers aspects of SDN and virtual middlebox networking to achieve the vision of SDCs. Baset et al.~\cite{baset:14} discussed how the concepts that compose SDCs can be leveraged to achieve efficient fault recovering and enhanced understanding of what constitutes regular system operation---a feature authors call ``operational excellence''. The proposed architecture focuses on high level features that enable operational excellence, and therefore it complements the features proposed by our architecture that enable the realization of the concept of SDCs.


Recent research on system virtualization focused on optimizing the technology for cloud data centers, in order to improve its security~\cite{zhang:11}, or providing scalable management systems for the VMs in the data center~\cite{nurmi:08}. Network virtualization has been extensively studied to augment the standard network technologies stack, which is hard to modify~\cite{chowdhury:10}. Chowdhury and Boutaba~\cite{chowdhury:10} present an extensive survey in the area. More recently, a survey by Jain and Paul~\cite{jain:13} focused on the challenges of network virtualization and SDNs in the specific context of cloud computing. Koponen et al.~ \cite{koponen:14} presented a system enabling network virtualization in multi-tenant data centers such as cloud data centers.  The technology is based on the concept of \emph{network hypervisor} and it provides one of the possible building blocks for our proposed architecture. 

Software-Defined Networking~\cite{sdnreport} is a core emerging concept enabling SDCs. In this field, Koponen et al.~ \cite{koponen:10} proposed a system, called Onix, that operates as a control platform in large-scale data centers. Monsanto et al.~\cite{monsanto:13} proposed an approach enabling composition of SDNs. This can be seen as an important step towards component-based SDNs and SDN elements, which in turn will enable reuse of infrastructures and easier description of networks by users (which will be able to use ``network templates'' to describe their networking requirements).

Regardless the specific approach for realization of virtual networking in a cloud data center (network hypervisors or SDNs), the problem of mapping computing and network elements to physical resources, as well as mapping virtual links into physical paths, needs to be addressed. A survey on this problem---\emph{Virtual Network Embedding (VNE)}---has been proposed by Fisher at al.~\cite{fischer:13}.

Another emerging concept that combines features of SDNs and system virtualization is virtual middlebox virtualization~\cite{gember:12}. Sherry et al.~\cite{sherry:12} proposes that middleboxes operations are outsourced to the cloud. The proposed system, called APLOMB, was able to delegate most of middleboxes tasks to the cloud, although there is a small proportion of operations that cannot be effectively migrated or require utilization of Content Delivery Networks to operate efficiently. Qazi et al.~\cite{qazi:13} proposed SIMPLE, an approach that applies SDN to facilitate the management of middleboxes. Gember et al.~\cite{gember:14} proposed Stratos, which shares some of the objectives of SIMPLE but with extra features to better manage the dynamicity of cloud environments. Hwang et al.~\cite{hwang:14} proposed NetVM, an approach that combines middlebox virtualization and network virtualization to extend functionalities of SDNs.

Mobile cloud computing is an emerging research area. Honeybee~\cite{fernando:13a} is a framework to enable mobile devices to offload tasks, utilize resources from devices, and perform human-aided computations. Huerta-Canepa~\cite{canepa:10} proposed an architecture to offload computation to nearby devices  using P2P techniques. Flores and Srirama~\cite{flores:14} proposed an approach for mobile cloud by exploring a middleware component between the devices and the cloud. Chun et al.~\cite{chun:11} proposed an approach to offload parts of a computation to the cloud to reduce energy consumption on the device.


In relation to the problem of energy-efficient cloud computing, recent research investigated the utilization of migration and consolidation to this goal~\cite{tsai:12,geronimo:13,beloglazov:13}. These approaches disregard applications running on VMs, and thus they do not consider the impact of the consolidation and the migration on the performance of particular applications inside the VMs. Chetsa et al.~\cite{chetsa:12} developed a white box approach targeting HPC applications where applications' characteristics are inferred at runtime and measures for energy savings are applied based on the application characteristics.

In our previous research, we presented the initial exploration of many of the issues discussed in this paper, including mobile cloud computing~\cite{hwang:14a,flores:14a}, energy-efficient cloud computing~\cite{beloglazov:13,beloglazov:13a,salehi:12,beloglazov:XX}, and dynamic provisioning and scheduling~\cite{mattess:10,calheiros:12a,calheiros:14}. This previous research constitutes the building blocks for more advanced aspects not considered in previous work, including: (i) how to utilize heterogeneous VM types for more efficient provisioning and application scheduling (as the previous work is based on homogeneous VM types or address unrelated programming models such as workflows); (ii) support for emerging programming models such as stream programming; and (iii) how to leverage software-defined networks to enable software defined clouds.

\section{System Requirements and Application Models} \label{sec:scenario}

Cloud data centers are composed of thousands of servers and hundreds of switches connecting the servers. Each server can host up to tens of virtual machines, which can belong to multiple customers. Depending on the applications running on the VMs, they can have different requirements in terms of CPU, network, and storage access.  VMs belonging to a single user can be organized on one or more VLANs  to better accommodate applications' demands.

As the number of customers increase, the management of all these aspects inevitably becomes complex. A software-defined cloud enables the configuration of the underlying infrastructure to be managed by a layer that coordinates the often conflicting configuration needs of different requests for cloud resources. 

In particular, the identified requirements for a system enabling Software-Defined Cloud Computing are:

\begin{itemize}
\item Support for high level description of user requirements and SLAs in terms of performance and needs of the mobile application,  computing platform, and the network;
\item \emph{Scalability} to support multiple simultaneous users with conflicting requirements and a large number of physical resources;
\item Capacity to quickly and dynamically modify previous configurations of the infrastructure to accommodate new demands;
\item Efficient utilization of cloud data center resources, achieving the required SLAs with the minimum possible resource utilization;
\item Efficient utilization of electricity to operate the infrastructure, so that all the aforementioned requirements are met with the highest energy efficiency. The highest energy efficiency requires a heavy \emph{consolidation} of both computing and network resources; and
\item Efficient support for mobile cloud applications. This means that a balance needs to be achieved between energy expenditure in the device and in the data center, under certain resource constraints. The balancing task is challenging because many factors need to be considered, such as amount of data transfer, mobile's energy  and resources utilization incurred by each possible decision.
\end{itemize}

\subsection{Application Models}

There are many types of applications that can benefit from the proposed infrastructure. In this section, we discuss some of them in more details.

\subsubsection*{Independent applications}

The most trivial application of distributed architectures for computation are \emph{embarrassingly parallel applications}. This class of applications are composed of tasks that execute independently from the others, and thus the tasks that compose the application can execute in any order and the failure of one task does not affect the others. A well-known application model that follows this pattern is \emph{Bag of Tasks} (BoT) applications. Tasks in BoT are completely independent, and thus can execute even a different binary file. Execution time between tasks can vary significantly, especially when tasks execute different code. \emph{Parameter sweep} is another application model that falls in this category. In this model, the application is constituted by the execution of the same binary subject to different input parameters on each task. Thus, it tends to be more homogeneous than BoT. Aneka's Thread and Task programming models~\cite{calheiros:12} follow a similar execution paradigm, however they differ from conventional parameter sweep by providing also a programming environment where software developers only need to describe the application logic, while all the aspects of the application that handle distribution and execution of tasks in the platform are abstracted away from developers.

\subsubsection*{MapReduce}

As large scale data centers with thousands of hosts became popular, new programming models that can take advantage of them arose. MapReduce~\cite{dean:08} is one such paradigms. A MapReduce job is composed of two types of tasks: Mappers and Reducers. Mappers receive data as input and generate intermediate (key, value) pairs that are consumed by Reducers, which generate the output as a result. MapReduce is becoming the programming model of choice for Big Data analytics, and thus one can expect that the popularity of this programming model will increase as the popularity of Big Data analytics increases.

\subsubsection*{Stream and IoT applications}

Another class of applications that are rapidly growing is streaming-related applications. The challenge here is being able to process a large, continuing stream of data. Sources of streams can be social networks (for example, Twits), or data from sensors, such as surveillance cameras. Given that the volume of data generated by these sources can be huge, it cannot be kept in the working storage area that is available for query processing. Therefore, decision is needed, as the streams arrive, on whether it should be processed or stored~\cite{rajaraman:11}. Also, because of the context in which streams are applied, along with the need for fast answers, approximate answer to queries related to streaming tend to be acceptable~\cite{rajaraman:11}. Streaming processing is strongly related to the emerging IoT paradigm~\cite{gubbi:13}, where there is a constant flow of information between sensors, web sites, and other sources and consumers of information.

\subsubsection*{Web applications}

One of the most prominent application classes in cloud computing is multi-tier web applications. These applications are usually composed of a presentation tier, which in clouds is handled by web servers and is responsible for returning http content to users; an application tier, which implements the business logic of the application and is processed by application servers or containers; and a database tier that stores persistent data used by the application. Well-engineered web applications allow each layer to scale independently to accommodate user demand. Unlike previous models that are usually batch processing activities, user requests for web applications tend to be short-lived, should be answered in a fraction of second, and tend to present bursty behavior.

\subsubsection*{Mobile applications}

The growing interest for smartphones also motivated the development of new applications for these devices. Mobile applications should be able to outsource large computations to the cloud, reducing the energy consumption of the device. Mobile applications also can make use of multiple cloud services to built ``rich applications''. However, a challenging aspect of mobile applications is the fact that mobile devices may not be all the times in an area under coverage of networks; in this case, the application must be able to undertake on the device itself some of the activities that otherwise would be carried out on the cloud. 

\section{System Architecture} \label{sec:architecture}

\begin{figure*}[!t]
\centering
\includegraphics[width=1.6\columnwidth]{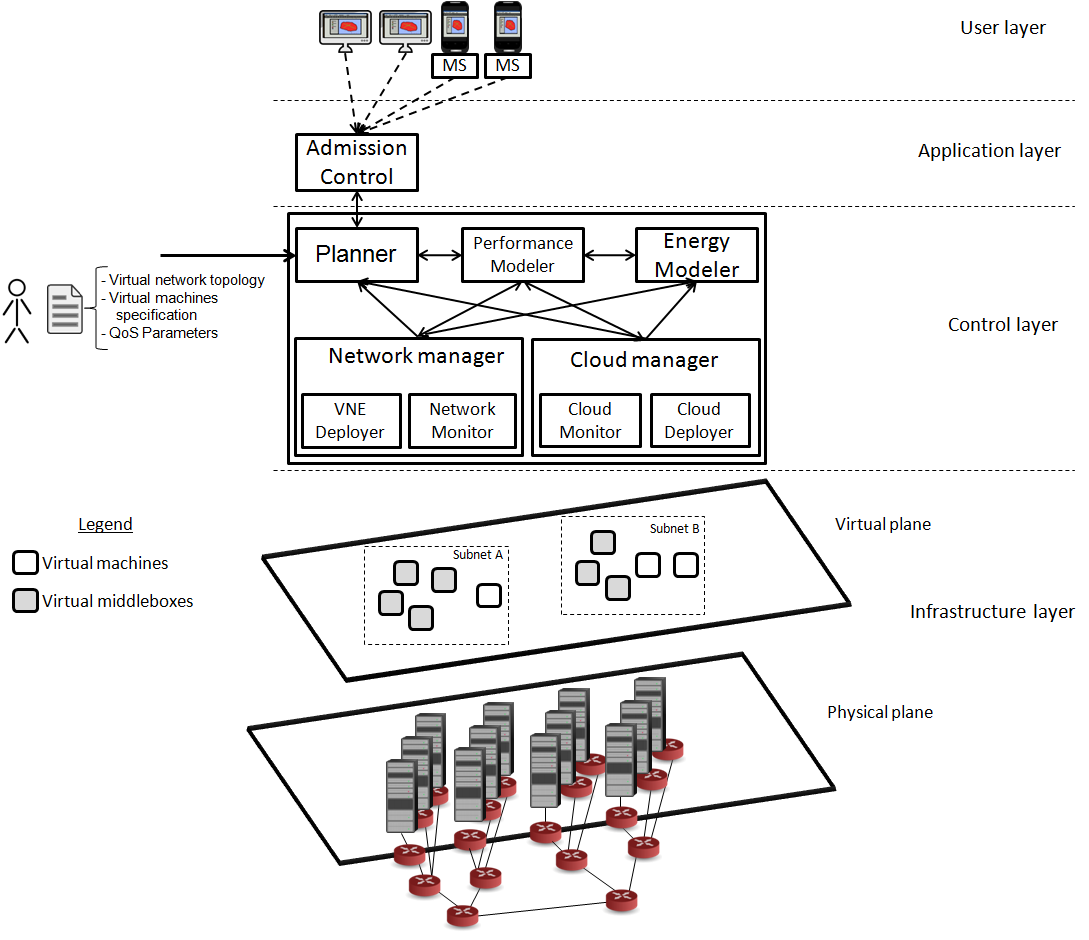}
\caption{An architecture for Software-Defined Cloud Computing environments.}
\label{fig:overview}
\end{figure*}

To enable a cost-efficient realization of user-defined virtual infrastructures in the cloud, we propose an architecture for Software-Defined Cloud Computing environments that is composed of four distinct layers, as depicted in Figure~\ref{fig:overview}.

The first layer, \emph{user layer}, runs on user devices, such as mobile devices and browsers from workstations. It provides an interface between the end user and the resources on the cloud, forwarding requests to the latter to complete certain tasks that can be better completed in the cloud rather than in the device itself. The second layer, the \emph{application layer}, is the level that decides whether requests can be executed or not and also schedules them. The next layer is the \emph{control layer}, where the logic that controls the SDC cloud is implemented. The bottommost layer, the \emph{infrastructure layer}, is the portion where the management actions from the layer above are applied, generating two distinct views: the the physical plane, which contains the physical resources that compose the data center, and the virtual plane, where the virtual infrastructure defined by users are realized.

In the rest of this section, we discuss each component of the proposed architecture.

\paragraph{Mobile Scheduler (MS)}

The Mobile Scheduler operates on mobile devices from users. It decides which activities from the request will be performed in the device and which activities will be delegated to the cloud. Such a decision is based on two key factors: capacity of the end-user device and energy-efficiency. The scheduler also submits to the cloud the tasks it decided to outsource.

\paragraph{Admission Control}
On the top layer, user requests are received by the cloud infrastructure by the \emph{Admission Control} component. This component decides whether the request can be accepted or not based on a number of factors such as user's identity (whether the user is authorized to use application), users' credit, load of the system, etc. 

\paragraph{Planner}

The Planner is the component responsible for deploying the SDC on behalf of the customer. It receives the requirements and decides the placement of the elements in the infrastructure. It includes the following activities:

\begin{itemize}
\item Deciding the host where each user virtual machine will be deployed;
\item Deciding in which host each virtual middlebox will be deployed;
\item Deciding the virtual routes connecting the VMs and virtual middleboxes.
\end{itemize}

\paragraph{Performance Modeler}

This module applies previous knowledge about the infrastructure and the current resource usage from running VMs to present an estimate of what is going to affect the performance of hosts and the network upon a particular VM placement decision made by the Planner.

\paragraph{Energy Modeler}

This module implements an energy model that provides to the Planner information about the current energy usage of the infrastructure and about the impact that particular placement decisions will have on the energy-efficiency of the infrastructure. 

\paragraph{Network Manager}

The Network Manager executes the different networking operations necessary to realize the virtual networks calculated by the Planner. The activities performed by this component include bandwidth reservation, subnetwork configuration, definition of routes in the network elements, and so on. It also monitors the status of the network to ensure that the required behavior is being achieved and maintained.

The allocation of physical paths and bandwidth to virtual routes is known as \emph{Virtual Network Embedding} (VNE)~ \cite{fischer:13}. It consists in mapping a series of graphs (representing the virtual topologies) into a larger graph representing the physical topology. This problem is complex because it requires many aspects to be considered, in particular:

\begin{itemize}
\item When composing the physical path, the aggregated latency added by each hop cannot exceed the maximum latency required by the virtual link. Thus, each virtual network will have a maximum number of switches that can compose the virtual link;

\item The maximum bandwidth of a virtual link is limited to the smallest bandwidth available among the physical links that compose the virtual network;

\item As new requests for embeds arrive in the system, previous embeds might need to be reconfigured to make room for the new requests;

\item As the VNE is an NP-Hard problem~\cite{fischer:13}, optimal solutions for large instances of the problem (as found in cloud data centers) are infeasible. Therefore, heuristic solutions, able to find a suitable solution in reasonable amount of time, is required for the problem;

\item Current solutions for the VNE problem ignore the problem of energy-efficient embedding. Thus, energy-efficient algorithms for the VNE problem need to be developed.
\end{itemize}

To achieve the above, the Network Manager implements two subcomponents. The \emph{Network Monitor} checks the status of all the network elements and links and reports the results to the Planner, so the Planner can consider this information when making decisions (e.g, avoiding routing traffic through a specific switch if it seems to be malfunctioning). The second component, the \emph{VNE Deployer}, is responsible for the realization of the VNE. To achieve this, it interacts with third-party SDN control software to expose the infrastructure request or interacts directly with the infrastructure layer using standard control plane interfaces such as OpenFlow~\cite{mckeown:08}.

Besides the networking aspects of SDC, the overall composition of the SDC also needs to look at the computing part of the problem, i.e, the mapping of VMs and virtual middleboxes to hosts. This is the task of the Computing Manager component of the architecture, which is discussed next.

\paragraph{Computing Manager}

The Computing Manager is the counter part of the Network Manager that is responsible for the realization of VM management operations on the physical infrastructure. Tasks from this module include creation, destruction, and migration of virtual machines, as well as their reconfiguration if required. This module also monitors the status of the computing infrastructure (physical and virtual) to ensure the required behavior is achieved and that there are no problems with the physical infrastructure.

Formally, the problem of mapping VMs to physical hosts can be modeled as a bin packing problem, where the resources required by all the virtual resources (machine or middlebox) assigned to a host cannot exceed its capacity. As in the bin packing problem, the challenge is how to build the mapping that minimizes the number of hosts (bins) in order to reduce the energy consumption. However, other challenges not faced in traditional bin packing problems arise:

\begin{itemize}
\item The application performance within virtual machines should be taken into account whenever it is known. This is because the performance of the hosted system can be degraded if there are interferences from other VMs on the same host. In particular, VMs that are known to demand the same type of resources should be mapped to different hosts to avoid contention for resources such as I/O and cache;

\item Virtual middleboxes tend to be CPU-intensive, and they are not necessarily defined as a fixed amount of resources to be allocated, as in the case of VMs allocated on users behalf;

\item The bin packing problem is an NP-Hard problem. In the context of Software-Defined Cloud Computing, it needs to be solved together with the VNE problem, which is also an NP-Hard problem. Thus, automatic deployment of SDCs is a very complex problem that needs to be solved in a reasonable amount of time for large instances (as data centers commonly contain thousands of physical resources) and potentially for many requests.
\end{itemize}

The Computing Manager is also composed of two subcomponents. The \emph{Cloud Monitor} checks and processes the status and performance of all the computing elements and virtual machines and reports relevant information back to the Planner. The second component, the \emph{Cloud Deployer}, manages the virtual machines and virtual middleboxes by creating, destroying, and migrating them as required. This requires interaction with the particular cloud management software used in the data center, or alternatively via direct interaction with hypervisors.

\section{Evaluation} \label{sec:evaluation}

The initial evaluation of the proposed framework is conducted via simulation. We utilize the CloudSim toolkit~\cite{calheiros:11}. CloudSim is a simulator of cloud computing environments that allows the description of data centers, as well as users workloads, to enable evaluation of new scheduling and provisioning policies in terms of energy efficiency, performance of hosted applications, and cost. The simulator has been extended to support modeling and simulation of SDCs.

We start the section by detailing the extensions we developed for modeling of SDCs, and then we present two use cases for SDCs. The first use case addresses the issue of differentiated services via QoS-aware allocation of bandwidth for requests and the second use case addresses bandwidth-aware VM placement. 

\subsection{CloudSim Extensions for SDCs}

\begin{figure*}[!t]
\centering
\includegraphics[width=.8\linewidth]{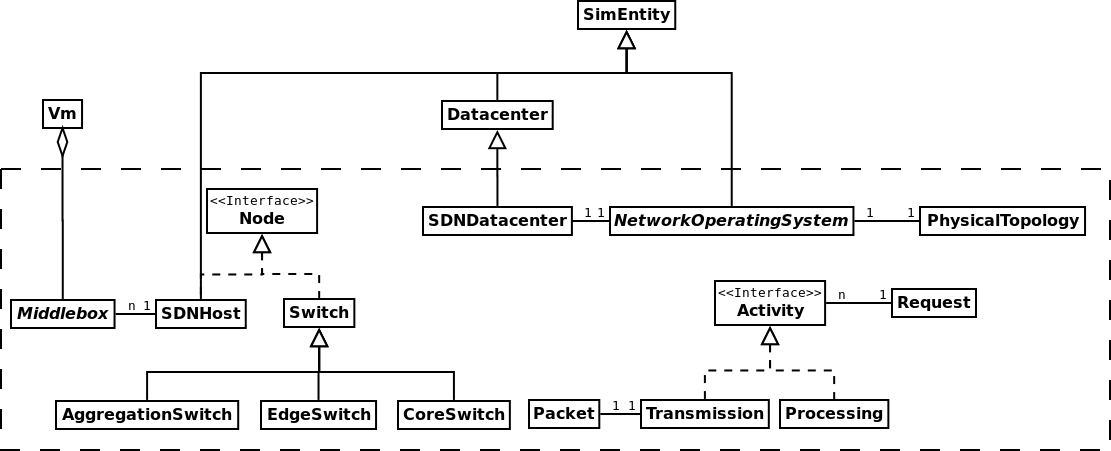}
\caption{Class diagram of the proposed CloudSim extension for simulation of Software-Defined Clouds.}
\label{fig:class}
\end{figure*}

The class diagram of the main classes of the SDC-related extensions of CloudSim is shown in Figure \ref{fig:class}. Firstly, new classes for modeling of the network infrastructure were designed\footnote{Although some of these classes have the same name as in some classes of NetworkCloudSim~\cite{garg:13}, they are different classes that lie in different Java packages---org.cloudbus.cloudsim.network.datacenter  and org.cloudbus.cloudsim.sdn.} to represent the different types of switches present in data centers~\cite{alfares:08}. The infrastructure layer also contains an SDNHost object, which is able to send and receive packets and forward them to the VMs and middleboxes running on the host. Both SDNHost and Switch implement a Node interface that contains methods that allow routing to be specified. 

The NetworkOperatingSystem class coordinates resource sharing at the network level. It creates virtual channels that represent a QoS guarantee for an application in terms of allocated bandwidth for a specific application and the maximum latency between VMs running the application. It also configures the routing tables of all the classes that implement the Node interface. The PhysicalTopology is associated to the NetworkOperatingSystem and contains a description of the infrastructure (its switches, hosts, and network links). The NetworkOperatingSystem is associated to an SDNDatacenter that extends CloudSim's Datacenter.

Middleboxes are modeled as a composition of one VM and extra attributes. The operation is modeled as a short-duration CPU intensive computation that, on completion, modifies the underlying request (by changing packet's sender and/or destination and/or size). This is an abstract class that needs to be extended to implement policies that represent the operation of specific middleboxes functions (e.g., load balancing, NAT, etc).

A user request (in the class Request) is modeled as a sequence of activities (modeled via an interface Activity) that can be either computation (class Processing) or communication (class Transmission) that model respectively the execution of a user request on the host and the flow of communication between different servers (e.g., a request to a database server originated from the web server's execution of a http request received from the user).

Both the physical infrastructure (hosts, switches, and the links between them) and the virtual topologies (VMs, virtual middleboxes, and the required connectivity between them) are supplied to the simulator via JSON files.

Finally, all the network elements in the new SDN extension support dynamic routing, whereas routing was statically-defined when elements were created in previous versions. This ensures that network simulations in CloudSim conform to the expectations of software-defined networks. 

\subsection{Use Case 1: QoS-Aware Bandwidth Allocation}

\begin{figure}[!t]
\centering
\includegraphics[width=.9\columnwidth]{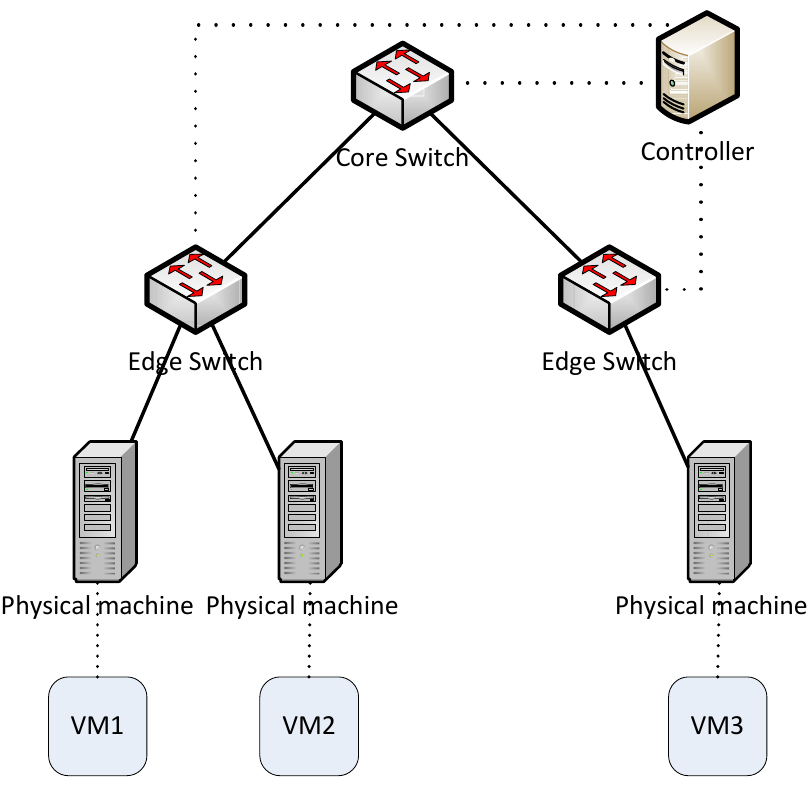}
\caption{Physical topology of the simulated data center utilized in the experiments.}
\label{fig_topo}
\end{figure}

The main purpose of this experiment is to show how SDCs (specifically controllers in SDN) can be used to serve users with different QoS requirements (response time). For this use case, we consider an environment with the physical topology depicted in Figure~\ref{fig_topo}. It comprises a data center with 3 hosts via a fat-tree topology with 2 edge switches and 1 core switch. Each edge switch connects two physical hosts as its leaves. A virtual environment is deployed to host a 3-tier web application (front end web server, application server, database server), which was modeled following the model proposed by Ersoz et al.~\cite{ersoz:07}. As each VM is configured to utilize the full resources of a physical machine, each physical machine hosts only one VM. In this way, we can ignore impacts of VM placement policies on final results. At the same time, this configuration provides flexibility of simulating various network traffic conditions, as some network routes use only edge switches, while others may use core and edge switches. 

In the simulation environment, via a controller, we can create separate virtual channels for different data flows. The idea is to allow  priority traffic to consume more bandwidth than normal traffic. For evaluation purposes, virtual channels between VMs are dynamically segmented to two different  channels, priority channel and standard channel. By default (when SDC is not used) a standard channel is used to transfer data between VMs where the bandwidth is evenly shared among all packets in the same channel. In contrast, a specific amount of bandwidth is exclusively and dynamically allocated for the priority channel, and thus such a bandwidth becomes unavailable for other channels.

\begin{figure}[!t]
\centering
\includegraphics[width=.8\columnwidth]{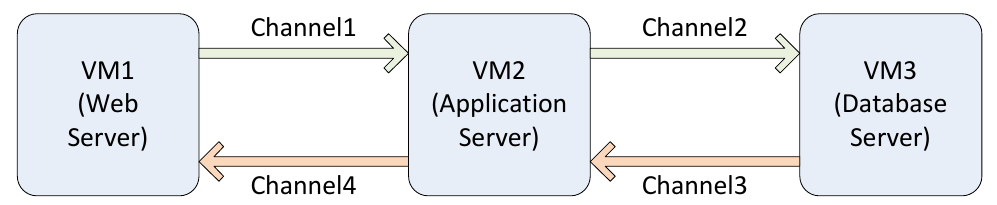}
\caption{Virtual channels between VMs.}
\label{fig_channel}
\end{figure}

Figure~\ref{fig_channel} shows the virtual channels between VMs. Channel 1 and Channel 2 are configured as standard channels, whereas Channel 3 and Channel 4 are set up first as standard and then as priority channel.

Network traffic is generated synthetically based on a typical web application model~\cite{ersoz:07}. Table~\ref{tab_data} shows the characteristics of the data used for the evaluation. Each request has a submission time and a list of processing and transmission descriptions. Requests are modeled as a sequence of activities from the web server to application server and then to the database server and all the way back to the web server. For each request, we measure response time to verify whether we can improve the QoS for the priority requests.

\begin{table}[!t]
\centering
\renewcommand{\arraystretch}{1.3}
\caption{Characteristics of requests used for evaluation. Requests are based on the model proposed by Ersoz et al.~ \cite{ersoz:07}.}
\label{tab_data}
\centering
\begin{tabular}{p{1.7cm}|c|c}
\hline
 & Distribution & Parameters\\
\hline\hline
Request inter-arrival times & Log-normal Dist. & $\mu$=1.5627, $\sigma$=1.5458 \\
\hline
Packet sizes & Log-normal Dist. & $\mu$=5.6129, $\sigma$=0.1343 (Ch1)\\
& &  $\mu$=4.6455, $\sigma$=0.8013 (Ch2)\\
& &  $\mu$=3.6839, $\sigma$=0.8261 (Ch3)\\
& &  $\mu$=7.0104, $\sigma$=0.8481 (Ch4)\\
\hline
Workload sizes & Pareto Dist. & location=12.3486, shape=0.9713\\
\hline
\end{tabular}
\end{table}

We consider three different network congestion levels: low, medium and high. For each case, requests for the normal traffic were sent at different rates to the standard channel: 100, 250 and 500 requests per second for low, medium, and high congestion respectively. At the same time, priority requests were sent at a fixed rate, 100 requests per second. 

\begin{figure}
\centering
\includegraphics[width=.95\columnwidth]{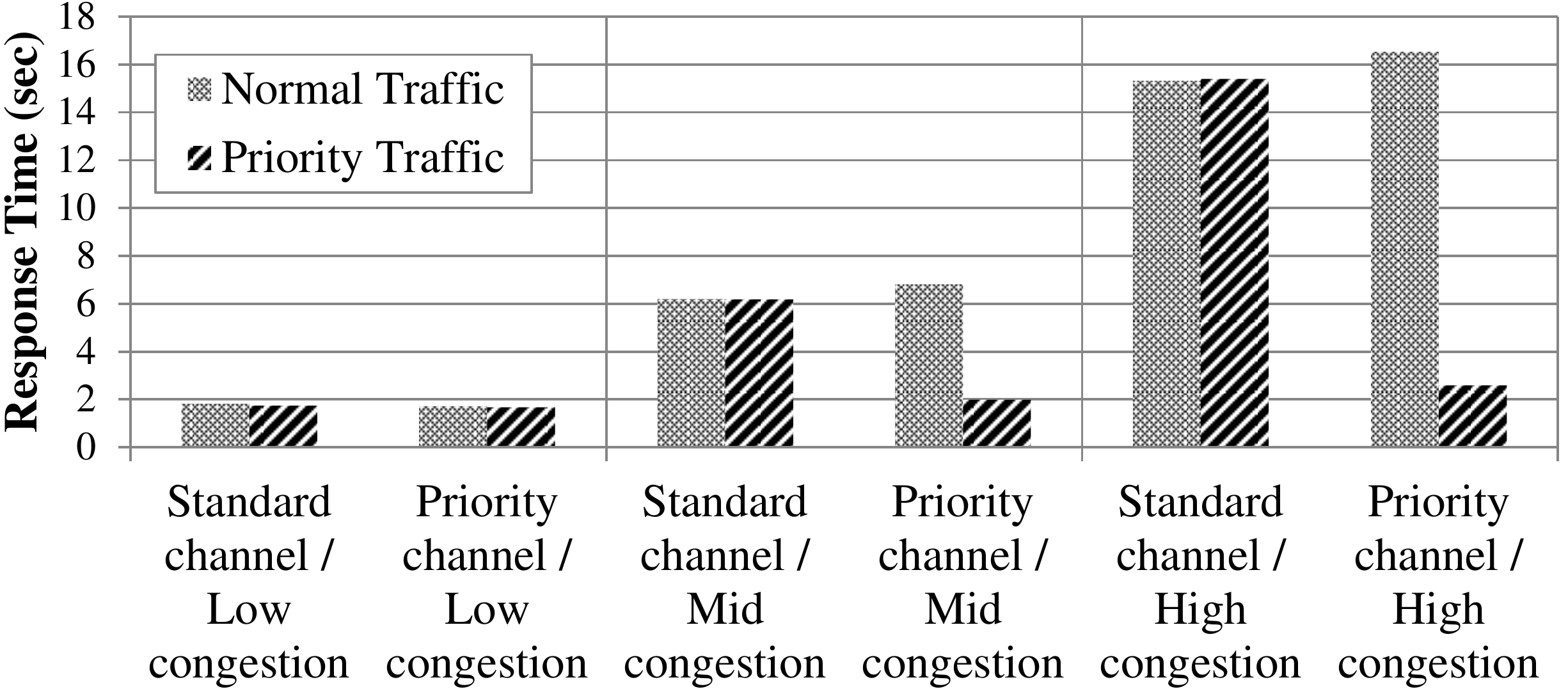}
\caption{Average response time of normal and priority requests for different network traffics}
\label{fig_res10}
\end{figure}

Figure~\ref{fig_res10} shows average response times for normal and priority requests under different network conditions. For all the cases, initially one channel is shared between all the requests and then  SDC is used to dynamically allocate a priority channel for priority requests. In the case of low congestion, response times of both normal traffic and priority requests are close when priority requests used the standard channel and  priority channel. However, when requests are sent at higher rates, the traffic using a priority channel had a significant improvement on the response time. For example, in the medium congestion case with 250 requests sent per second as normal traffic, the average response time for priority requests decreased from 6.176 sec to 1.990 seconds when priority channel is available for priority traffic. Similarly, when the normal requests are sent at the rate of 500 requests per second on the standard channel, the priority channel enabled priority requests to be served within 2.589 seconds, while for the priority traffic that uses a standard channel response time increased to 15.389 seconds. 

With this use case, we show how cloud providers can use SDC to offer services with various QoS levels. As demonstrated, there is a certain amount of bandwidth reserved for the priority channel that allows priority requests to be served in much shorter time. Although response times for the normal traffic became slightly longer, almost constant response time can be obtained for requests using priority channel. Using this feature of QoS, providers can assure higher QoS for customers that pay a higher cost for receiving better levels of service (e.g. a paid version of mobile application) or have more critical requirements.

\subsection{Use Case 2: Network-Aware VM Placement}

Our second set of experiments aim at showing how SDCs can help in reducing energy consumption of cloud infrastructures by enabling consolidation of VMs during placement, enabling network elements connecting unused hosts to be shut down. In this case, we consider both general-purpose VMs and virtual middleboxes. We simulated a data center with 40 hosts. Each host has 16 cores, each core with capacity of 4000 MIPS,  connected to a Gigabit switch. 

We generated 100 VM creation requests where the types of VMs are randomly selected among the configurations described on Table~\ref{tab:usecase2}. Each request also has a different start time and a life time following exponential distribution and Pareto distributions respectively~\cite{mills:11}, thus the VM is destroyed after the set life time expires. As VM placement is a bin packing problem, we compared two widely utilized heuristics for this purpose, namely Best Fit and Worst Fit. Because the packing problem described in this use case has two dimensions (bandwidth and CPU power), normalized requirements in terms of CPU (product between number of cores and required power per core) and bandwidth (in terms of the host capacity) are used. The Best Fit algorithm selects the most utilized host with enough resource to serve the VM request, so VMs can be consolidated in the smaller number of hosts. On the other hand, the Worst Fit algorithm selects the least loaded host to maximize computation power. For both cases, idleness is calculated based on the available area of the 2-dimensional space that represents available resources. 

\begin{table*}
\centering
\caption{Virtual Machine configurations used for the VM consolidation experiments.}
\label{tab:usecase2}
\begin{tabular}{c||c|c|c|c|c} \hline
VM Type		&	Web Server  	&	App Server & DB Server	& Proxy			& Firewall		\\ \hline\hline
Mips				& 2000				& 3000			& 2400			& 2000			& 3000 			\\
Cores			& 2						& 8					& 8					& 8					& 8					\\
Bandwidth	& 100 Mbps			& 100 Mbps		& 100 Mbps		& 500 Mbps		& 500 Mbps		\\ \hline
\end{tabular}
\end{table*}

Table~\ref{tab_vmplace} shows the results for this experiment. Energy consumption is calculated based on the linear power model presented by Pelley et al.~\cite{pelley:09}. As the result shows, the data center can save up to 26\% of its energy consumption by using SDC techniques along with the Best Fit algorithm for placing a VM. Moreover, at any given time a maximum of 25 hosts are in use, which means 15 hosts are idle throughout the whole experiment. Those 15 hosts can be turned off to save more energy. SDNs also enable network devices to be turned off by the controller~\cite{jin:13}. Thus, when network elements only connect  those 15 off-lined hosts, they could be turned off to save energy, which results in further savings in energy consumption.

\section{Challenges and Opportunities} \label{sec:challenges}

Achieving the vision of Software-Defined Cloud Computing requires extensive research and development in key topics, discussed in the rest of this section. As the area of SDC matures and the topics below start being addressed,  more research questions are likely to be identified.

\subsection{Mobile Cloud Computing}

There are important open questions in the area of Mobile Cloud computing that need to be addressed. The first one concerns the decision of where mobile applications should be executed. First, the whole computation could be outsourced to the cloud to reduce the energy consumption of mobile devices. A second approach concerns deployment of \emph{cloudlets} that are placed between the devices and data centers. They provide an approach with smaller communication latency for the outsourced computation. Computation can also be split and performed partially on the device and partially on the data center or cloudlet. Each decision has implications in terms of latency, performance, and energy consumption. Finally, another approach for outsourcing computation is distributing computation among nearby mobile devices that form an \emph{ad hoc} network of volunteer devices, a concept that utilizes elements from established research in the area of volunteer computing~\cite{anderson:03}. 

Another open problem is a programming model optimized for mobile applications. Appropriate programming models can help in generating more efficient partitioning of applications, and can accelerate the adoption of the technology by providing standard frameworks that could be incorporated to currently available platform-specific mobile application development environments.

\subsection{Accurate Modeling and Prediction of the Infrastructure's Energy Consumption and Performance}

One important aspect of achieving SDCs is the ability to answer what-if questions about potential virtual resource placement and its impact on  energy consumption. This will enable the Planner to optimize the placement and minimize resource and energy wastage to operate the infrastructure.

To achieve the above goals, research towards accurate modeling of energy consumptions by applications, virtual resources, and physical resources is required. The model should be able to operate both as a ``black box'' model of virtual resources (i.e., without knowledge about the applications hosted by the user's virtual infrastructure) and ``grey box'' model (where some knowledge about the application is available). The need for supporting both models goes beyond more accurate energy modeling: because techniques for reduction of energy consumption can impact application performance. More aggressive energy reduction techniques could be applied if one can be confident about the real performance impact on the application.

In parallel with the previous need for energy consumption modeling, the Planner needs to be able to estimate the performance impact that its placement decisions can incur on the system. The same discussion about black or gray boxes modeling also applies in this case.

\begin{table}
\centering
\renewcommand{\arraystretch}{1.3}
\caption{Accumulated energy consumed by physical hosts and the maximum number of simultaneously utilized hosts for different VM placement algorithms.}
\label{tab_vmplace}
\centering
\begin{tabular}{c|c|c} \hline
Algorithm	& Energy consumption by hosts (Wh) & Max Hosts \\ \hline \hline
Best Fit &	164025 &	25 \\ \hline
Worst Fit	& 221841 & 40 \\ \hline
\end{tabular}
\end{table}

\subsection{Market and Pricing Models for SDCs}

Better market and pricing models than ``one size fits all'' are necessary to enable SDCs. For example, the provider could offer better prices if the user request could be slightly adapted to enable more efficient and/or balanced use of physical resources. Incentives could also be present to help in driving workloads out of peak time to times where there are more free resources. This could be achieved with the application of negotiation techniques where requests from users would be replied with counter-offers from providers proposing one or more potential different configurations and the discount price to be applied to each of the options. This change in configuration could be on ``time'' (i.e., delaying the request to a point in the future) or on ``space'' (i.e., changing the specification of the required virtual infrastructure). Users could accept one of the suggestions of the providers or go ahead with the initial requested configuration at the provider's full price.

Moreover, because knowledge about the user workload can help in increasing the resource and energy efficiency, providers of SDCs should offer incentives for users that disclose information about the general type of applications to be executed in the infrastructure (e.g, CPU-intensive, I/O-intensive, parallel, MapReduce, and so on).

\subsection{Combined VNE and VM Placement Problems}

As discussed previously, both problems of finding an optimal virtual network embed for the virtual links from user's requests and the problem of optimal location of virtual resources (both virtual machines and virtual middleboxes) are NP-Hard problems. Therefore, to be solved in feasible time, heuristics and meta-heuristics need to be applied.

More research is necessary in order to understand better the interplay between these two placement problems and to determine how the goal of energy- and resource-efficient solution to the problems can be achieved while performance of applications is guaranteed.

\subsection{Energy-Efficient Middlebox Virtualization}

Most of the research in energy-efficient cloud computing has so far focused on reducing energy usage of CPUs for general purpose VMs. Virtual middleboxes provide an extra challenge in the field as they are CPU-intensive VMs whose performance can affect the speed of the network traffic and thus impact the performance of applications in unpredictable ways. How to reduce energy consumption of virtual middleboxes without impacting the network traffic is an open research challenge.

\subsection{Language for Description of SDCs}

Research on how to provide to users intuitive and simple ways of describing all their requirements are needed. In this sense, focus could be on ontologies for description of SDCs or on Domain Specific Languages (DSLs) that capture all the intricacies of an SDC.

Furthermore, user requirements could be described in different abstraction levels, and simultaneous support for all of them is required in order to expand the user base of SDCs. Ideally, users could range from engineers with clear idea of the virtual infrastructure they want (number and types of VMs and network topology and middleboxes between them) to managers with a high level view of the system demands.

The language should also cater for the elasticity of the cloud, so simpler ways to describe how the system should scale up and down with the demand must be part of the solution.

\subsection{Autonomic Cloud Computing}

Finally, the view of Software-Defined Cloud Computing can only be fully implemented with research and development towards \emph{autonomic cloud computing}. The first reason for this is that cloud data centers are composed of thousands of hardware elements that can fail at any time. Thus, to enforce SLAs, failures of components need to be detected as early as possible (self-management) and corrective steps needs to be automatically performed in order to speed up the recovery (self-healing). Second, clouds are elastic per nature, and the demand of users, as well as number of SDC requests, can vary along the time. The Planner needs to react to changes in number and types of requests to achieve the system's goals of resource and energy efficiency (self-optimization).

Finally, security of all these processes is an intrinsic part of the system for users to trust the SDC provider and utilize it (self-protection).

\section{Conclusions} \label{sec:conclusion}

Software-Defined Cloud Computing is emerging as a result of advances in the areas of cloud computing, system virtualization, software-defined networks, software-defined middleboxes networking, and network virtualization. Before SDCs become a reality, however, many challenges need to be overcome.

In this paper, we presented an architecture enabling SDCs focusing on variety of applications including compute and data-intensive applications in Web, mobile, and enterprise environments. We discussed the different elements that comprise the architecture and evaluated  through simulation the potential of SDCs in two use cases---QoS-aware bandwidth allocation and bandwidth-aware, energy-efficient VM placement. We also discussed the open challenges and opportunities arising from this emerging area.

As SDCs and the enabling technologies progress, we expect new challenges to arise and new application scenarios to emerge that will make SDCs a mainstream technology with applications in all the industry sectors.

\end{document}